\documentclass[12pt]{article}
\usepackage{graphicx}

\newcommand{\beq}{\begin{equation}}
\newcommand{\eeq}{\end{equation}}
\newcommand{\bea}{\begin{eqnarray}}
\newcommand{\eea}{\end{eqnarray}}

\newcommand{\gsim}{\lower.7ex\hbox{$
\;\stackrel{\textstyle>}{\sim}\;$}}
\newcommand{\lsim}{\lower.7ex\hbox{$
\;\stackrel{\textstyle<}{\sim}\;$}}

\def\ot{{\bf T}}
\def\cp{{\bf CP}}
\def\cpt{{\bf CPT}}

\begin{document}

\thispagestyle{empty}
\vspace*{-10mm}

\begin{flushright}
UND-HEP-05-BIG\hspace*{.08em}04\\
hep-ph/0509153 \\
\today \\
\end{flushright}
\vspace*{4mm}

\begin{center} 
{\large \bf Hadronization -- the Unsung Hero} \\
{\large \bf rather than the Alleged Villain in the Tale of 
\cp~Violation \footnote{Invited Talk given at {\em Hadron 05}, Rio de Janeiro, August 22 -- 26, 2005} }
\end{center}
\vskip 0.3cm \centerline{I. I. Bigi$^a$} 

\vskip 0.3cm 
\centerline{$^a$Dept. of Physics, University of Notre Dame du Lac, Notre Dame, IN 46556, U.S.A.}
\centerline{e-mail: ibigi@nd.edu}
\vskip 1.0cm

\centerline{\bf Abstract}
The novel successes scored by the Standard Model of High Energy Physics in the 
last few years concerning heavy flavour dynamics do not weaken the case for `New Physics' 
around the TeV scale. They do suggest however that one cannot 
{\em count} on that New Physics impacting heavy flavour decays in a numerically massive 
way. Yet studying this impact will be essential in diagnosing the features of the New Physics. In particular the decays of beauty hadrons have to be analyzed with considerable precision on the experimental as well 
as theoretical side. While hadronization effects often represent the main bottleneck 
in our understanding in the short run, they will provide powerful and discriminating tools in the long run, when applied comprehensively and judiciously. The expertise required to exhaust the discovery potential in $B$ decays does exist in the hadron physics community or can be developed without needing a new breakthrough -- yet a greater effort has to be made to communicate it to the heavy flavour community. 



\section{Prologue}

There are three reasons why I am grateful for the invitation to participate in this conference and speak to you. Terry Goldman expressed the third reason the other day: Rio de Janeiro gives a new and very pleasant meaning to the word `winter conference'. The second reason will be explained in my talk. My 
first reason is of a general nature. I am experiencing high energy physics as a truly noble human activity and view it as the most global one, where I cannot see any drawback 
to globalization -- contrary to other manifestations of globalization. However as far as its geographic 
distribution is concerned, I am exaggerating: marking on a globe where high energy physics research is done, you will find it happens almost exclusively  in the northern hemisphere. This lack of balance concerns me, and I hope for a wider distribution in the future. We owe a lot of gratitude  to our colleagues, who maintain and further build 
high energy physics in the southern hemisphere. I view CBPF here in Rio as the flagship of fundamental physics in 
the southern hemisphere, appreciate the work Alberto Reis and his colleagues are doing in `spreading 
the gospel' and promise my support to them.

\section{Introduction}

Around the turn of the millenium -- in the years 1998 -- 2001 -- a `quantum jump' in our knowledge 
(though {\em not} our understanding) of fundamental physics has occurred: 
\begin{itemize}
\item 
"the completion of an era": the establishment of {\em direct} \cp~violation in $K_L$ decays; 
\item 
"the beginning of a new journey": the first observation of \cp~violation in a system other than  
the $K^0 - \bar K^0$ complex, namely in  $B_d \to \psi K_S$;  
\item
"messages from the heavens": the discovery of neutrino oscillations; 
\item 
"the birth of a Standard Model of Cosmology": a consistent picture concerning the composition of 
our Universe 
\footnote{If indeed there exist other  `universes' besides ours, the term {\em uni}verse is no longer appropriate. With our `universe' being just one among countless others, also the name `cosmos' is misplaced. For this Greek word refers to an order that derives its beauty from careful arrangement.} 
in terms of visible and dark matter and `dark energy' and the distribution of the cosmic microwave radiation and its fluctuations has emerged. The evidence for `dark energy' 
can best be commented by repeating Rabi's famous quote about the muon:"Who ordered that?"

\end{itemize}
The first three items certainly and the fourth one probably impact our views of microscopic dynamics 
profoundly, although it is still too early to say how. Concerning the first two items I would like to 
emphasize without an explanation (that can be found in my Varenna lectures \cite{VARENNA}) 
that the SM has scored not `merely' more or even new successes, but actually novel ones, i.e. 
of a qualitatively new kind: the {\em predicted} `Paradigm of large \cp~violation in $B$ decays' 
\cite{BS80} 
has been verified through the discovery of indirect and direct \cp~violation in 
$B_d \to \psi K_S$, $B \to K\pi$ and presumably also in $B_d \to \pi^+ \pi^-$ 
\cite{CPASYMDISC,CPASYMEXP};  
these large \cp~asymmetries are commensurate with \ot~violation 
with{\em out} assuming the full power of 
\cpt~invariance \cite{TM02}.  

{\em Yet none of these novel successes of the SM weaken the case for it being incomplete and for 
New Physics being even `nearby', i.e. around the TeV scale.} 
We have `heavenly' data showing the incompleteness of the SM: baryogenesis, neutrino 
oscillations, `dark matter' and the most puzzling feature of all, `dark energy'. 
On the theoretical side we still have not solved the Strong CP Problem \cite{CPBOOK}. 

More generally these  successes do not shed light on any of the mysterious features of the SM. 

\noindent {\bf (i)}  
What are the dynamics driving the electroweak symmetry breaking of 
$SU(2)_L\times U(1) \to U(1)_{QED}$. How can we tame  the instability of Higgs dynamics with its quadratic mass divergence? 
I find the arguments compelling that point to New Physics at the 
$\sim 1$ TeV scale -- like low-energy SUSY; therefore I call it the `confidently predicted' New Physics 
or {\em cpNP}. 

\noindent {\bf (ii)}  
The family structure with the quantization of the electric charges of quarks and leptons as expressed 
for example through 
\beq 
Q_e = 3Q_d
\eeq 
This would be naturally explained through Grand Unification at very high energy scales 
implemented through, e.g., $SO(10)$ gauge dynamics. 
I call this the `guaranteed New Physics' or {\em gNP}. 

\noindent {\bf (iii)}  
{\em Finite family replication}: we infer from the observed width of $Z^0$ decays that there are  
three (light) neutrino species. The hierarchical pattern of CKM parameters as revealed by the data is so peculiar as to suggest that some other dynamical 
layer has to underlie it. I refer to it as `strongly suspected New Physics' or {\em ssNP}. Saying we pin our hopes for explaining the family replication on Super-String or M theory is a scholarly way of saying 
we have hardly a clue what that {\em ssNP} is. 
 
Comprehensive studies of heavy flavour dynamics might provide insights into 
the {\em gNP} and {\em ssNP}, items {\bf (ii)} and {\bf (iii)}, but I would not count on it. 
{\em However -- and this is the central element of my message -- they will be crucial in identifying the {\em cpNP}. } 

In a nutshell my message reads as follows: 
We expect confidently that New Physics surfaces at the TeV scale. 
Yet we have to aim beyond `merely' establishing the existence of New Physics -- our goal has to be to identify its salient features.  
TeV scale dynamics is likely to have some nontrivial 
impact on $B$ decays, and the   
discovery potential in $B$ decays is essential for figuring out the {\em cpNP}, {\em not} a 
luxury. 
Yet due to the past `unlikely' success of the CKM description one can{\em not count} on 
{\em massive} manifestations of New Physics, at least not in $B$ decays typically. 
Therefore we need high accuracy both on the experimental and theoretical side in heavy flavour 
studies. 
This requires a better {\em quantitative} understanding of {\em hadronization} to exhaust 
the discovery potential in $B$ decays. The necessary expertise exists or can be acquired 
with{\em out} a new theoretical breakthrough. 

By the past `unlikely' success of the CKM description I mean the following. As far as we know large fractions of the observables $\Delta M_K$, $\epsilon_K$ and $\Delta M_B$ could be 
due to physics beyond the SM and even most of $\epsilon ^{\prime}$. Likewise the constraints from the 
data on the CKM parameters translate into seemingly broad bands in plots of the 
CKM unitarity triangle \cite{FITTER}. 
The problem with this statement is that it is not even wrong -- it just misses the main point: 
it is already very remarkable and highly nontrivial that all these constraints can be incorporated in a 
meaningful way in such plots. This becomes more obvious when plotting these and other observables 
from the strange, charm and beauty transitions on a commensurate energy scale. They cover several orders of magnitude between $10^{-8}$ and $10^{-16}$ MeV. That the CKM expectations are always within 50 \% or so of the data over this range is most remarkable and could not be expected a priori -- in particular when coupled with the fact that these predictions are based on values for the CKM and other mass parameters that would seem frivolous at best -- if they were not forced upon us by independent measurements. The huge top quark mass of about 175 GeV (which 
is the only quark mass not only to approach the vector boson masses, but  even to exceed them considerably) provides a case in point. There could easily have been inconsistencies. In other words:  
for CKM theory to provide a satisfactory description of the data, its parameters had to reside in a
very peculiar corner of the general parameter space -- yet nature has put them exactly into 
that `neighbourhood'   
\cite{CPBOOK}. Therefore some of us had been rather confident that the \cp~asymmetry measured in 
$B_d \to \psi K_S$ would be close to the CKM prediction, whatever the latter would be 
-- as indeed was the case. 

The observation of this asymmetry in 2001 by both BABAR and BELLE  \cite{CPASYMEXP} 
has established the 
CKM paradigm as a {\em tested} theory that no longer deserves the patronizing label of an 
`ansatz'. Furthermore  a `demystification' of \cp~violation has occurred: we have seen that if the dynamics are sufficiently rich as to be able to support \cp~violation, the latter can be large (although 
we still do not know how to give meaning to the notion of `maximal' \cp~violation). This process of 
demystification will have been completed, if \cp~violation is found anywhere in the lepton sector -- a point I will return to later. We also know that the standard CKM theory is utterly irrelevant for 
baryogenesis, that the observed baryon number of the Universe shows the need for New Physics 
\cite{DOLGOVVAR}. 

In summary: we have to conduct dedicated and comprehensive searches for New Physics. You have heard that before, have you not?  And it reminds you of a quote by Samuel Beckett: 
\begin{center}
"Ever tried? Ever failed? \\
No matter. \\
Try again. Fail again. Fail better."
\end{center}
Only an Irishman can express profound skepticism concerning the world in such a poetic way. 
Beckett actually spent most of his life in Paris, since Parisians like to listen to someone expressing such a world view, even while they do not share it. Being in the service of Notre Dame du Lac, the home of the `Fighting Irish', I cannot just ignore such advice. Yet my response is: "Cheer up -- we know there is New Physics (see above) -- we will not fail forever!" 

My friend Antonio Masiero likes to say: "You have to be lucky to find New Physics." True enough -- yet 
let me quote someone who just missed by one year being a fellow countryman of Masiero, namely 
Napoleon, who said: "Being lucky is part of the job description for generals." Quite seriously I think 
that if you as  
an high energy physicist do not believe that someday somewhere you will be a general -- maybe not 
in a major encounter, but at least in a skirmish -- then you are in the wrong line of business. 

In the following I will focus on \cp~violation for rather pragmatic reasons; they are one of the most 
sensitive probes for New Dynamics, since \cp~asymmetries can be {\em linear} in a New Physics amplitude rather than quadratic and thus exhibit an enhanced sensitivity to New Physics. 

The remainder of my talk will be organized as follows:  after singing the praise of 
hadronization in Sect.\ref{PREL} I will sketch `King Kong' scenarios for New Physics searches 
in Sect.\ref{KONG}; then I will return to the need for precision and illustrate it with lessons from 
the extraction of $V(cb)$ in Sect.\ref{VCB} and case studies concerning hadronization in 
Sect.\ref{ALLY}. 
Sect. \ref{APPEAL} contains an outlook and a personal appeal. 

\section{Prelude: Singing the Praise of Hadronization}
\label{PREL}

Hadronization and nonperturbative dynamics in general are usually viewed as unwelcome 
complication, if not outright nuisances. A case in point was already mentioned: while 
I view the CKM predictions for $\Delta M_K$, $\Delta M_B$, $\epsilon_K$ to be in 
remarkable agreement with the data, significant contributions from New Physics could 
be hiding there behind the theoretical uncertainties. While this is factually correct, it misses a deeper 
truth. {\em Without} hadronization bound states of quarks and antiquarks will not form; without 
the existence of kaons $K^0 - \bar K^0$ oscillations  obviously cannot occur. 
It is hadronization that provides the `cooling' of the (anti)quark degrees of freedom, which 
allows subtle quantum mechanical effects to add up coherently over macroscopic distances. 
Otherwise  
one would not have access to a super-tiny energy difference Im${\cal M}_{12} \sim 10^{-8}$ eV, 
which is very sensitive to different layers of dynamics, 
and indirect \cp~violation could not manifest itself. The same would hold for $B$ mesons and 
$B^0 - \bar B^0$ oscillations. 

To express it in a more down to earth way:  
\begin{itemize}
\item 
Hadronization leads to the formation of kaons and pions with masses exceeding 
greatly (current) quark masses.  
It is the {\em hadronic} phase space that suppresses the \cp~{\em conserving} rate for 
$K_L \to 3 \pi$ by a factor $\sim 500$, since the $K_L$ barely resides above the three pion threshold. 
\item 
It awards `patience'; i.e. one can `wait' for a pure $K_L$ beam to emerge after starting out with a 
beam consisting of $K^0$ and $\bar K^0$. 
\item 
It enables \cp~violation to emerge in the {\em existence} of a reaction, namely 
$K_L \to 2 \pi$ rather than an asymmetry; this greatly facilitates its observation. 
\end{itemize}
For these reasons alone we should praise hadronization as the hero in the tale of \cp~violation 
rather than the villain it is all too often portrayed. 

\section{`King Kong Scenarios' for New Physics Searches}
\label{KONG}

This scenario can be portrayed as follows: "It is unlikely that one will encounter King Kong; yet 
once it happens there will be no doubt that one has come across something wildly out of the ordinary." 
The point of analogy is the following. When, say, a certain transition has been observed to proceed although it is predicted not to -- like $K_L \to 2\pi$ in 1964 -- or when predicted and 
observed rates differ by orders of magnitude, we can speak of a `qualitative' discrepancy, which establishes the existence of New Physics right away, though not its nature. This has happened with the decays 
of strange hadron -- and it might happen again. 

\subsection{\cp~Violation in Charm Decays}
\label{CPCHARM}

I can be very brief here, since a detailed exposition has been given in the `Cicerone' \cite{CIC} and we have heard a nice talk by D. Asner \cite{ASN}. 

The relative dullness of the weak SM phenomenology for charm can be harnessed as a laboratory for studying hadronization. Yet charm decays still have the potential to reveal New Physics as well. 
One should note that New Physics in general and flavour changing neutral currents in particular can affect down-type -- $d$, $s$ and $b$ -- and up-type quarks -- $u$, $c$ and $t$ -- quite differently; 
charm is the {\em only} up-type quark allowing the full range of probes for New Physics: 
\begin{itemize}
\item 
top quarks do {\em not} hadronize; 
\item 
$\pi ^0 - \pi ^0$ oscillations are not possible, since neutral pions are their own antiparticle; with so few 
decay modes for pions 
\cp~asymmetries are basically ruled out by \cpt~ symmetry. 
\end{itemize}
My basic contention is: charm transitions provide unique portals for obtaining novel access to 
New Physics with the experimental situation being a priori favourable (apart from the fact that its leading 
decays are Cabibbo allowed). Searches for \cp~violation constitute the most powerful and promising tool in the long run for several reasons: (i) {\em Baryogenesis} requires new sources of \cp~violation. 
(ii) {\em Within the SM}  the weak phase is highly diluted in once Cabibbo suppressed  modes, namely 
${\cal O}({\rm sin}^4 \theta _C)\sim 10^{-3}$,  and zero in Cabibbo favoured and 
doubly suppressed modes 
(except for $D^{\pm} \to K_S\pi^{\pm}$ \cite{YAMA}). 
(iii) A \cp~asymmetry can be {\em linear} in the New Physics amplitude and thus is more sensitive to it. 
(iv) Final state interactions are very active in the charm region. (v) The branching ratios for 
\cp~eigenstates are large. (vi) There is one fly in the ointment, though: $D^0 - \bar D^0$ oscillations 
are slow at best \cite{CIC}. 

$B$-factories clearly can make a tremendous contributions to \cp~studies in the charm sector. The 
challenge in particular to LHCb is: "Can you?" To which degree can LHCb search for an 
asymmetry in $D^{*+} \to D^0(t) +\pi^+ \to [K^+ \pi^-]_D+\pi^+$ vs. 
$D^{*-} \to \bar D^0(t) +\pi^- \to [K^- \pi^+]_D+\pi^-$ 
as a function  of the time of decay $t$?

\subsection{\cp~Violation in Leptodynamics}
\label{CPLEPTO}

There is a compelling impetus for searching for \cp~violation in leptodynamics, namely to complete 
the aforementioned `demystification' of \cp~violation and to get a firmer handle on baryogenesis 
as driven by primary leptogenesis \cite{LEPTO}. 

\subsubsection{Neutrino Oscillations}
\label{NUOSC}

When searching for a \cp~asymmetry in neutrino oscillations, you do not have to worry about hadronization. Yet one has to disentangle genuine \cp~violation from enhancements due to the environment being made up of matter rather than antimatter. This will likely turn out to be a challenging  
task. I would not be surprised if our colleagues engaged in this enterprise will rue previous 
requests to be free of hadronization. This reminds me of an ancient Greek saying: "When the gods 
really want to harm you, they fulfill your wishes." 
 
\subsubsection{Electric Dipole Moments}
\label{EDM}

Truly impressive experimental sensitivity has been achieved concerning the electric dipole moments 
of electrons and neutrons:
\beq
d_e = (0.07 \pm 0.07) \cdot 10^{-26} \; {\rm ecm} \; , \; d_N < 0.63 \cdot 10^{-25} \; {\rm ecm}
\eeq
Even so they are still many orders of magnitude above CKM predictions 
\beq 
d_e^{CKM} < 10^{-36} \; {\rm ecm} \; , \; d_N^{CKM} <  10^{-30} \; {\rm ecm}
\eeq
On the other hand typical benchmarks for New Physics scenarios are 
\beq 
d_e ^{NP}\; , \;  d_N^{NP} \sim (10^{-28} - 10^{-26}) \; {\rm ecm} 
\eeq
a range that should be reached in the foreseeable future. 

\subsubsection{\cp~Violation in $\tau$ Decays -- the Next Hero Candidate}
\label{TAUHERO}

The most promising channels for exhibiting \cp~asymmetries are $\tau \to \nu K \pi$, since due to 
the heaviness of the lepton and quark flavours they are most sensitive to nonminimal Higgs dynamics,  
and they can show asymmetries also in the final state distributions rather than integrated rates 
\cite{KUHN}.  

There is also a {\em unique}  opportunity in $e^+e^- \to \tau ^+ \tau ^-$: since the $\tau$ pair is produced with its spins aligned, the decay of one $\tau$ can `tag' the spin of the other $\tau$. I.e., 
one can probe {\em spin-dependent} \cp~asymmetries with {\em unpolarized} beams. This provides 
higher sensitivity and more control over systematic uncertainties. 

I feel these features are not sufficiently 
appreciated even by proponents of Super-B factories. It has been recently pointed \cite{BSTAU}  
out that based on known physics one can actually predict a 
\cp~asymmetry: 
\beq 
\frac{\Gamma(\tau^+\to K_S \pi^+ \overline \nu)-\Gamma(\tau^-\to K_S \pi^- \nu)}
{\Gamma(\tau^+\to K_S \pi^+ \overline \nu)+\Gamma(\tau^-\to K_S \pi^- \nu)}= 
(3.27 \pm 0.12)\times 10^{-3}
\label{CPKS}
\eeq
due to $K_S$'s preference for antimatter.

\section{The Need for Precision -- Extracting $V(cb)$ as a Lesson}
\label{VCB}

New Physics scenarios typically will not 
create massive deviations from the SM predictions in $B$ decays. Thus precision is essential on the experimental as well as theoretical side. This is not a noble, yet unreachable goal. It can be achieved when 
one combines a {\em robust} theoretical framework with comprehensive and detailed data. I will 
briefly illustrate it with a mature example, namely the extraction of $V(cb)$ from inclusive 
semileptonic $B$ decays. 

The robust theoretical framework there was provided by heavy quark expansions, which treat 
{\em non}perturbative effects through an expansion in inverse powers of the heavy quark $m_Q$.  
It provides an analytical algorithm genuinely based on QCD and receiving support from various 
sum rules; in the future it can incorporate findings from lattice QCD as well \cite{HQTH}. 

The total semileptonic width of $B$ mesons as well as the lepton spectra encoded through 
lepton energy and hadronic mass 
{\em moments} can be  expressed through heavy quark parameters 
$V(cb)$, beauty and charm quark masses $m_{c,b}$, and $B$ meson expectation values 
of local heavy quark operators \cite{IMPREC,URIGAMB}. 
With a mere handful of such parameters one can describe a host of observables. Those 
provide a large number of {\em overconstraints}. Obtaining a consistent fit is thus far from assured a priori. Achieving it provides an impressive demonstration that even systematic uncertainties are under control \cite{DELPHIVCB,BABARMOM}
. As an extra bonus one has found that the fitted values of these heavy quark parameters 
obey certain theoretical constraints although those were {\em not} imposed on the fit; furthermore they even agree with other determinations of these parameters. 
For example the (kinetic) $b$ quark mass 
can be determined in $B$ production just above threshold -- $\Upsilon (4S) \to b \bar b$ -- which is determined by the {\em strong and electromagnetic} forces, and in the {\em weak} decays of $B$ hadrons: 
\beq 
m_b^{kin}(1\; {\rm GeV})|_{\Upsilon (4S) \to b \bar b} = 4.57 \pm 0.08 \; {\rm GeV} \;  
vs. \; m_b^{kin}(1\; {\rm GeV})|_{B \to l \nu X_c} = 4.61 \pm 0.068 \; {\rm GeV}
\eeq 
With them being fully consistent, it makes sense to average them. 
A comprehensive analysis \cite{BUCH} yields (with the relative error given in paranthesis) 
\bea 
\nonumber 
m_b^{kin}(1\; {\rm GeV}) &=& (4.59 \pm 0.04) \; {\rm GeV} \; \; (\sim 1 \%)\\
\nonumber 
m_c^{kin}(1\; {\rm GeV}) &=& (1.14 \pm 0.06) \; {\rm GeV}   \; \; (\sim 5 \%)\\
|V(cb)| &=& (41.58 \pm 0.67) \cdot 10^{-3}  \; \; (\sim  1.6 \%)
\label{HENNING}
\eea
One should note here that the present uncertainty on $|V(us)|$ is about 1.1 \%. The numbers 
in Eq.(\ref{HENNING}) show that percent level precision is not utopian even when nonperturbative 
dynamics is involved.

\section{Case Studies for Hadronization as a Difficult Ally}
\label{ALLY}

$\Delta M_K$, $\epsilon _K$ and $\epsilon ^{\prime}$ -- weak observables of central importance -- 
are affected 
by hadronization effects like $\pi \pi$ phase shifts, the $\eta - \eta ^{\prime}$ wave functions, 
the role of the $\sigma$ resonance etc., yet in a way that has never been reliably quantified. 
Here I will list four case studies of a more 
recent provenance, which concern $B$ decays and where hadronization can actually be employed 
as a powerful tool -- if applied judiciously. 

First I would like to remind you that spectroscopy has been the subject of philosophical 
reflections over the centuries, the fruits of which can be distilled into the `three Razors of 
Spectroscopy': 
\begin{center} 
Ockham's Razor: \\
"Entities should not be multiplied unnecessarily!"\\
\vspace{3mm}
Peter Minkowski's Razor: \\
"No experiment can make two discoveries in the same data set!"\\
\vspace{3mm}
Stefano Bianco's Razor:\\
"You can learn a lot by cutting judiciously!" 
\end{center}
These rules are not necessarily binding; however you can ignore them only 
at your own peril. The first Razor is almost self-evident, although open to interpretation, in 
particular a posteriori. The second Razor can be based on many examples from history. 
For the third one, which is the only one with a positive content, I will give an example below. 

\subsection{Case I: The `1/2 $>$ 3/2 Puzzle'}

Beyond what charm spectroscopy can teach us about QCD there are (at least) 
three more motivations for understanding it: (i) To determine the total width for 
$B \to l \nu X_c$ and keep the error on it small, one needs accurate and realistic modeling of 
the hadronic system in the final state. (ii) One needs it to gauge the errors on the measurements 
of the exclusive modes 
$B \to l \nu D/D^*$ due to a feed down from higher resonances. 
(iii) With heavy quark symmetry decoupling the spin of the heavy quark, one can label heavy flavour hadrons by their total spin and the angular momentum $s_q$ carried by the light degrees of freedom 
\cite{SWANSON}. To the degree that charm can be treated as a heavy flavour, one has two 
meson groundstates -- $D$ \& $D^*$ -- and four excited states formed by P wave $c$$\bar q$ 
configurations, two of which carry $s_q = 3/2$ and are predicted to be narrow with the other two 
carrying $s_q = 1/2$ and being broad. 
With heavy quark theory one can derive sum rules from QCD  that relate heavy quark parameters -- 
quark masses, $B$ expectation values of heavy quark operators -- with subclasses of 
semileptonic $B$ decays, in particular $B\to l \nu D(s_q = 1/2 \, {\rm or} \, 3/2)$. If 
the P wave states saturate those sum rules -- a very natural, albeit not proven assumption -- then 
the production of the narrow $s_q = 3/2$ charm meson states should dominate over that of the 
broad $s_q = 1/2$ ones in semileptonic $B$ decays \cite{URIPUZZLE,CIC}. 
This prediction is in apparent conflict 
with DELPHI studies \cite{DELPHIVCB,BIANCO}. They 
show that the final states not being just $D$ or $D^*$ are made up more by 
broad than narrow charm states. One possible conjecture for resolving this discrepancy is that 
a significant fraction of the broad component is made up by {\em radial} excitations rather than 
P wave $D(s_q = 1/2)$. 

This is not an academic problem and actually goes beyond even precise determinations of $|V(cb)|$ 
and $|V(ub)|$: the SM predicts $b\to l \nu q$ to proceed by pure $(V-A)_q \times (V-A)_l$ 
currents. A $(V-A)_q \times (V+A)_l$ coupling is most unlikely even with New Physics, since the 
required right-handed neutrino is unlikely to be sufficiently light due to Majorana masses; yet  
New Physics could conceivably induce a $(V+A)_q \times (V-A)_l$ current. Its main impact would be 
on the {\em shape} of the lepton spectrum making it softer and thus changing 
its moments. However such an impact had to be 
disentangled from hadronization effects. Modeling incorrectly the charm final state could then either 
fake a signal for $(V+A)_q \times (V-A)_l$ currents or hide one.

\subsection{Case II: $\phi_1$ from $B_d \to 3$ Kaons} 

Analysing \cp~violation in $B_d \to \phi K_S$ decays is a most promising way to search 
for New Physics. For the underlying quark-level transition $b \to s \bar s s$ represents a pure 
loop-effect in the SM, it is described by a single $\Delta B=1$\& $\Delta I=0$ operator (a `Penguin'), a reliable 
SM prediction exists for it \cite{GROSS} -- sin$2\phi_1(B_d \to \psi K_S) \simeq {\rm sin}2\phi_1(B_d \to \phi K_S)$ -- 
and the $\phi$ meson represents a {\em narrow} resonance.  

Great excitement was created when BELLE reported a large discrepancy between the predicted and 
observed \cp~asymmetry in $B_d \to \phi K_S$. Based on more data taken, this discrepancy has 
shrunk considerably: the BABAR/BELLE average for 2005 yields \cite{CPASYMEXP}
\beq 
{\rm sin}2\phi_1(B_d \to \psi K_S) = 0.685 \pm 0.032
\eeq
compared to 
\beq 
{\rm sin}2\phi_1(B_d \to \phi K_S) 
\left\{
\begin{array}{ll} 0.50 \pm 0.25 ^{+0.07}_{-0.04} & {\rm BABAR}\\
0.44 \pm 0.27 \pm 0.05 & {\rm BELLE}
\end{array}
\right. \;  ; 
\eeq
BABAR's as well as BELLE's numbers are below the prediction, albeit by one sigma only. It 
is ironic that such a smaller deviation, although not significant, is actually more believable than 
the large one originally reported by BELLE. 

This issue has to be pursued with vigour, since this reaction provides such a natural portal to 
New Physics. One complication has to be studied, though, in particular if the observed 
value of sin$2\phi_1(B_d \to \phi K_S)$ falls below the predicted one by a moderate amount only. 
For one is actually observing $B_d \to K^+K^-K_S$. If there is a single weak phase like in the SM one finds 
\beq 
{\rm sin}2\phi_1(B_d \to \phi K_S) = - {\rm sin}2\phi_1(B_d \to `f_0(980)' K_S) \; , 
\eeq 
where $`f_0(980)'$ denotes any {\em scalar} $K^+K^-$ configuration with a mass close to that of the 
$\phi$, be it a resonance or not. A smallish pollution by such a $`f_0(980)' K_S$ -- by, say, 
10\% {\em in amplitude} --  
can thus reduce the asymmetry assigned to $B_d \to \phi K_S$ significantly -- by 20\% in this example. 

In the end it is therefore mandatory to perform a {\em full time dependent Dalitz plot analysis} 
for $B_d \to K^+K^-K_S$ and compare it with that for $B_d \to 3 K_S$ and 
$B^+ \to K^+K^-K^+, \, K^+K_SK_S$ and also with $D \to 3K$. This is a very challenging task, but 
in my view essential. There is no `royal' way to fundamental insights. 
\footnote{The ruler of a Greek city in southern Italy once approached the resident sage with the request 
to be educated in mathematics, but in a `royal way', since he was very busy with many 
obligations. Whereupon the sage replied with admirable candor:There is no royal way to mathematics.} 

An important intermediate step in this direction is given by one application of 
{\em Bianco's Razor}, namely to analyze the \cp~asymmetry in $B_d \to [K^+K^-]_MK_S$ as a 
function of the cut $M$ on the $K^+K^-$ mass.

\subsection{Case III: $\phi_2$ from $B_d \to $ {\em Pions}} 

Unlike the preceding case two different operators drive $B \to pions$, namely one obtained from 
a tree and one from the Penguin diagram. With only the first one depending on the weak phase $\phi_2$ it is highly nontrivial to extract its size from these decays, since the 
\cp~asymmetry depends on the relative weight of those two operators, which is shaped by 
strong dynamics over which we have less than full theoretical control. 

A theoretically clean way to disentangle the impact from the Penguin operator is to measure 
the isospin two final state $B \to [\pi\pi ]_{I=2}$ \cite{ITWO}. By measuring all rates 
$B_d \to \pi^+\pi^-, \, \pi^0 \pi^0$, $B^{\pm} \to \pi^{\pm}\pi^0$ this can be achieved in principle, 
but maybe not in practice with sufficient accuracy; $B_d \to \pi^0\pi^0$ provides the bottle neck. Less challenging experimentally are the modes $B \to \rho \pi$, but one pays a theoretical price there: 
the actual final state state consists of three pions, where the $\rho$ cannot be seen as a narrow resonance. Furthermore there are other contributions to the three-pion final state like 
$\sigma \pi$. It hardly matters actually whether the $\sigma$ is a bona fide resonance or some 
other dynamical enhancement. What matters is that merely cutting on the di-pion mass will not produce 
a $\rho \pi$ final state with sufficient purity; furthermore the $\sigma$ structure cannot be 
described adequately by a Breit-Wigner shape. As analyzed first in \cite{ANTONELLO} and then 
in more detail in \cite{ULF} ignoring such complications can induce a systematic uncertainty in the extracted value of $\phi_2$. The case of $B \to \rho \rho$, which experimentally is even better than the previous one, is even worse theoretically. We have to aim at a total uncertainty -- 
experimental plus theoretical -- of not exceding 5\%. 
To achieve this ambitious goal in $B \to 3 \pi$ we have to bring 
the full machinery of a time dependent Dalitz plot analysis to bear  augmented by our understanding of 
chiral dynamics.

\subsection{Case IV: $\phi_3$ from $B^{\pm} \to D^{neut}K^{\pm}$} 

As first mentioned in 1980 \cite{CS80}, then explained in more detail in 1985 \cite{BS85} 
and further developed in \cite{GRONWYL},  
the modes $B^{\pm} \to D_{neut}K^{\pm}$ should exhibit direct \cp~violation driven by the 
angle $\phi_3$, if the neutral $D$ mesons decay to final states that are {\em common} to 
$D^0$ and $\bar D^0$. Based on simplicity the original idea was to rely on two-body modes like 
$K_S\pi^0$, $K^+K^-$, $\pi^+\pi^-$, $K^{\pm}\pi^{\mp}$. One drawback of that method are the small 
branching ratios and low efficiencies. 

A new method was pioneered by BELLE and then implemented also by BABAR, namely to employ 
$D_{neut} \to K_S \pi^+\pi^-$ and perform a full Dalitz plot analysis. This requires a very 
considerable analysis effort -- yet once this initial investment has been made, it will pay handsome profit in the long run. For obtaining at least a decent description of the full Dalitz plot population 
provides  considerable cross checks concerning systematic uncertainties and thus a high degree of 
confidence in the results. BELLE and BABAR find \cite{CPASYMEXP}:  
\beq 
\phi_3 = 
\left\{
\begin{array}{ll} 68^o  \pm 15^o (stat) \pm 13^o(syst) \pm 11^o ({\rm model}) & {\rm BELLE}\\
70^o  \pm 31^o (stat) \pm 12^o(syst) \pm 14^o ({\rm model}) & {\rm BABAR}
\end{array}
\right.
\eeq 
I view it still a pilot study, yet a most promising one. It exemplifies how the complexities of 
hadronization can be harnessed to establish confidence in the accuracy of our results. I consider 
this to be the way of the future. 

\section{On HEP's Future Landscape -- a Call to Action \& an Appeal}
\label{APPEAL}

To me there is persuasive evidence of a theoretical as well as experimental nature that the SM is incomplete pointing to New Physics driving the electroweak phase transition: the 
confidently predicted New Physics {\em cpNP} at the TeV scale. This is the justification -- an excellent one in my view -- for the LHC. The goal has to go beyond establishing the presence of some New 
Physics -- we have to identify its salient features as well. One should keep in mind that SUSY -- 
for me the most likely candidate for the {\em cpNP} -- is not a true theory yet or even a class of theories: it is largely an organizing principle given our current lack of understanding its breaking. 

It has been recognized that the LHC is primarily a discovery machine sweeping out huge regions 
of `terra incognita'; this lead to the proposal of a linear collider as a more surgical and focussed probe 
for the {\em cpNP} -- yet the same justification applies to flavour factories! While 
comprehensive studies of heavy 
flavour transitions are not very likely to shed light on the {\em ssNP} behind the flavour 
puzzle of the SM (although they might), they will be essential in elucidating salient features 
of the {\em cpNP}. For New Physics at the TeV scale can affect flavour transitions 
significantly, and analyzing them is thus 
complementary to the program of the LHC and Linear Collider. It is actually 
essential to obtain all the information experimental research can give us 
on Nature's Grand Design. 
{\em Dedicated and comprehensive studies of heavy flavour dynamics are thus a necessity, 
not a luxury.}  

New Physics having a significant impact on $B$, $\tau$ and charm decays does not mean it will 
be numerically massive and thus obvious. We must succeed in adding the element of 
`high accuracy' to that of `high sensitivity'. 

I view a Super-B factory as crucial for us achieving that high accuracy and 
decoding the {\em cpNP}. It requires  
close collaboration between experiment and theory. A central 
message of my talk is to treat hadronization and nonperturbative dynamics in general as our ally -- albeit a complex and sometimes 
quirky one -- rather than as a nuissance. I would like to combine this with a personal appeal: The 
expertise required to attain an essential goal, namely to exhaust the discovery potential in heavy 
flavour transitions by harnessing low-energy hadronization does exist or can be acquired 
with {\em no} need for a breakthrough. However it tends to reside in a community all too often disjoint from the 
heavy flavour community -- {\em this has to change!} We need input from studies of 
$\pi \pi$, $K \bar K$, $K \pi$ etc. final state interactions at low energies, in nonleptonic as well as 
semileptonic charm decays and in $B \to multi$neutrals, and we have to refine and extend our understanding of chiral dynamics. These are very complex tasks; when tackling them we should remember 
the example of Swiss watches for guidance: those became famous by being reliable and sturdy, not necessarily 
elegant. 

One final remark: The pion has been discovered in 1947. Its dynamics have been studied 
intensely for the last sixty years. This conference here in magnificent Rio has shown 
that it is not a closed chapter; a great deal still has to be learnt. This demonstrates much better than 
many words, how momentous the discovery of the pion was, 
how profound a paradigm shift it created. We thus owe a 
great deal of gratitude to our late colleague Cesar Lattes for his work, as summarized in the following Memorial.

\vspace{0.5cm}

{\bf Acknowledgments:}  I am grateful to Alberto Reis and his colleagues for organizing such an enjoyable meeting. Being able to stage it in Rio with its breathtaking vistas gave them huge advantage, 
of course. 
This work was supported by the NSF under grant PHY03-55098.


\end{document}